\documentclass[journal]{IEEEtran}
\usepackage[cmex10]{amsmath}
\usepackage{algorithm}
\usepackage{algpseudocode}
\usepackage{amssymb}
\usepackage{array}
\usepackage{balance}
\usepackage{bm}
\usepackage{booktabs}
\usepackage{color}
\usepackage{cite}
\usepackage{epsf}
\usepackage{epsfig}
\usepackage{graphicx}
\usepackage{hyperref}
\usepackage{multirow}
\usepackage{stfloats}
\usepackage{subfigure}
\usepackage{tabularx}
\usepackage{url}
\usepackage{verbatim}
\ifCLASSINFOpdf
\else
\fi

\begin{document}
% paper title
\title{Data-Rate Driven Transmission Strategies for Deep Learning Based Communication Systems}

\author{
\IEEEauthorblockN{
        Xiao Chen,
        ~Julian Cheng, \IEEEmembership{Senior Member, IEEE},
        ~Zaichen Zhang, \IEEEmembership{Senior Member, IEEE},
        ~Liang Wu, \IEEEmembership{Member, IEEE},
        ~Jian Dang, \IEEEmembership{Member, IEEE},
        ~Jiangzhou Wang,~\IEEEmembership{Fellow,~IEEE}
        }
\thanks{Manuscript received March 30, 2019; revised August 01, November 30, 2019 and January 10, 2020; accepted January 13, 2020.
This work was supported by NSFC projects (61501109, 61571105, and 61601119), national key research and development plan (2016YFB0502202), Scientific Research Foundation of Graduate School of Southeast University (YBJJ1816), the Scholarship from China Scholarship Council (201806090072), and Zhishan Youth Scholar Program of SEU.
The editor coordinating the review of this paper and approving it for publication was V. Aggarwal.
(\emph{Corresponding authors: Zaichen Zhang; Liang Wu.})}
\thanks{X. Chen, Z. Zhang, L. Wu and J. Dang are with National Mobile Communications Research Laboratory, Southeast University, Nanjing,
210096, China (email: \{chen$\_$xiao, zczhang, wuliang, dangjian\}@seu.edu.cn).}
\thanks{J. Cheng is with School of Engineering, The University of British Columbia, Kelowna, V1V 1V7, BC, Canada (email: julian.cheng@ubc.ca).}
\thanks{J. Wang is with School of Engineering and Digital Arts, University of Kent, Canterbury, CT2 7NT, United Kingdom (email: J.Z.Wang@kent.ac.uk).}
}

\maketitle

\begin{abstract}
Deep learning (DL) based autoencoder is a promising architecture to implement end-to-end communication systems.
One fundamental problem of such systems is how to increase the transmission rate.
Two new schemes are proposed to address the limited data rate issue: adaptive transmission scheme and generalized data representation (GDR) scheme.
In the first scheme, an adaptive transmission is designed to select the transmission vectors for maximizing the data rate under different channel conditions.
The block error rate (BLER) of the first scheme is $80\%$ lower than that of the conventional one-hot vector scheme.
This implies that higher data rate can be achieved by the adaptive transmission scheme.
In the second scheme, the GDR replaces the conventional one-hot representation.
The GDR scheme can achieve higher data rate than the conventional one-hot vector scheme with comparable BLER performance.
For example, when the vector size is eight, the proposed GDR scheme can double the date rate of the one-hot vector scheme.
Besides, the joint scheme of the two proposed schemes can create further benefits.
The effect of signal-to-noise ratio (SNR) is analyzed for these DL-based communication systems.
Numerical results show that training the autoencoder using data set with various SNR values can attain robust BLER performance under different channel conditions.
\end{abstract}


\section{Introduction}
To satisfy growing demand for various communication applications and services, the next-generation network must deliver enhanced mobile broadband, ultra-reliable and low-latency communications, and massive Internet of Things (IoT) ecosystems \cite{6736746, 7414384, 7041045, 7736615}.
One primary concern is to accommodate the exponential rise in the number of user equipments and the traffic capacity in future communication systems.
Hence, several promising technologies have been proposed, and they include massive multi-input and multi-output (MIMO) transmissions, millimeter wave communications, ultra-dense networks, and non-orthogonal multiple access \cite{6415388, 6736761, 5783993, 7422408, 7811838, 7982784}.
For these conventional communication systems, there exist a number of limitations, such as unavailable channel state information in complex transmission scenario, high complexity to process big data, and sub-optimal performance caused by conventional block structure.
For these reasons, with the significant development of deep learning (DL) \cite{6817512, schmidhuber2015deep, 8382166}, researchers have applied the machine learning (ML), especially DL technologies, to design communication systems for benefits that cannot be obtained using the conventional approaches \cite{5670617, 7792374, 8233654, xiaohuai, huang2019incorporating}.

As a promising technique, deep learning implements communication systems using deep neural networks (NNs).
Different from the conventional communication system that consists of multiple independent blocks (e.g., source/channel coding, modulation, channel estimation, equalization), the DL-based communication system can jointly optimize transmitter and receiver for end-to-end performance without a block structure \cite{8054694, 8214233}.
DL-based system design is promising for the following reasons: (i) A DL-based communication system can be optimized for end-to-end performance by using deep NNs, which is fundamentally different from the block-structure in conventional communication systems;
(ii) A DL-based communication system can be optimized for a practical system over any type of channel without requiring a tractable mathematical model, and this includes the channel models that take into account of different transmission scenarios and non-linearities;
(iii) DL algorithms can provide faster processing speed than conventional communication algorithms, since the execution of NNs can be highly parallel on concurrent architectures and can be implemented using low-precision data types \cite{37631}.

Attracted by these advantages, there have been a number of studies on DL-based communications and signal processing using state-of-the-art tools and hardware \cite{7852251, 8242643, 7926071, 7886039, 8054694, 8422339, o2018over, 8214233, felix2018ofdm, 8322184, 8353153, 8482358, 01151, 8254356, 8437142, 8283585, 8052521, kim2018deepcode}.
The DL method is used to deal with certain challenges in existing communication systems.
For example, the DL-based belief propagation algorithm was originally used to improve the performances of channel decoding, where low-complexity and near optimal decoder performance were obtained \cite{7852251, 8242643, 7926071}.
Around the same time, autoencoder was developed to address the problem of learning an efficient physical layer \cite{7886039}.
In DL theory, an autoencoder describes a deep NN to find a low-dimensional representation of its input at certain intermediate layer that allows reconstruction at the output with minimal error \cite[Ch. 14]{Goodfellow-et-al-2016}.
The DL-based communication system can be represented and implemented by an autoencoder that is trained using the dataset offline.
Then, the trained autoencoder can be directly applied to practical systems online.
A DL-based communication system, interpreted as an autoencoder, performs an end-to-end reconstruction task that jointly optimizes transmitter and receiver as well as learns signal encoding \cite{7886039, 8054694, 8422339, 8437142}.
To address the challenges of frame synchronization,  an autoencoder was proposed to represented a complete communication system \cite{8214233, felix2018ofdm}, and comparable performance can be achieved even without extensive hyperparameter tuning.
More recently, a DL-based algorithm has been used to solve the channel state information feedback and channel estimation problems in massive MIMO systems, and it outperforms the state-of-the-art compressive sensing based algorithms \cite{8322184, 8353153, 8482358}.

For future communication systems, there is a huge demand for data rate due to an increasing number of communication devices and equipment types, and improved quality of services (QoS).
Consequently, high data-rate schemes should be developed in DL-based communication systems for future wireless networks.
However, one-hot vector \cite{176685}, being the most commonly used data representation in existing studies \cite{8054694, 8233654, 8214233, 8422339, o2018over, felix2018ofdm, 7926071, 8437142}, has a low data rate in DL-based communication systems.
The reason is that an $M\times 1$ one-hot vector consists of $0$s in all entries with the exception of a single $1$, e.g., $[0,\ldots,0,1,0,\ldots,0]^T$, and there are only $M$ possible transmitted messages.
Here, the value of $M$ cannot be large since the oversize $M$ will lead to prohibitive training complexity and time-consuming training for the autoencoder.
Hence, a small value of $M$ leads to limited data rate.
This becomes a barrier for developing future DL-based communication systems.
Besides, the autoencoder with one-hot vector is typically trained using a fixed vector size $M$, which becomes a constraint when designing communication systems having different data rate requirements.
Also, the conventional autoencoder is trained under a fixed signal-to-noise ratio (SNR) value with an unrealistic expectation to operate well for a wide range of SNR values in practical transmission scenarios \cite{7414384}.
It was reported that training the autoencoder at different SNR values will affect autoencoder performances \cite{8054694}, but there is no detailed study on the effect on such a system.
Therefore, our objective is to design a new transmission scheme and replace the conventional one-hot vector scheme to achieve higher data rate.
As well, we will investigate the effect of training SNR on the performance of DL-based communication systems.
Here, training SNR denotes the fixed SNR used for training the autoencoder offline, and it can be different from the practical SNR of a communication system when it is operating online.

In this paper, an adaptive transmission scheme is first designed for different communication scenarios to maximize the data rate in DL-based communication systems having a QoS constraint.
Then, we propose a generalized data representation (GDR) scheme to improve the data rate of DL-based communication systems.
Finally, we analyze the effect of SNR and mean squared error (MSE) performance in DL-based communication systems.
Comparable block error rate (BLER) performance can be achieved by the proposed transmission schemes which has lower complexity and higher data rate than the conventional DL-based communication system\footnote{Notably, throughout this paper, the conventional DL-based communication system refers to an autoencoder based communication system that adopts the one-hot vector data representation.}.

The major contributions of this paper are summarized as follows:
\begin{enumerate}
  \item In DL-based communication systems, we pointout the limited data rate problem of the conventional one-hot vector scheme.
      To address this issue, we design an adaptive transmission scheme having a QoS constraint for different channel conditions.
      In the proposed scheme, the optimal transmission vectors are adaptively selected for different SNR values, where the goal is to maximize the data rate with a constraint on MSE performance.
      It is shown that, when both two schemes have the same data rate, the proposed adaptive transmission scheme can reduce BLER of the conventional one-hot vector scheme by $80\%$.
  \item Furthermore, we propose a generalized data representation scheme to improve the data rate in DL-based communication systems.
      The proposed scheme represents the message by using a probability vector having multiple non-zero elements, instead of the conventional one-hot vector having only one non-zero element.
      As expected, higher data rate is obtained by the proposed GDR scheme with comparable BLER performance and low complexity.
      When the vector size is eight, as an example, the proposed GDR scheme can double the data rate of the conventional one-hot vector scheme.
      To the best of the authors' knowledge, this is the first time that the GDR scheme is proposed and its effectiveness is verified.
  \item We investigate the effect of SNR on the system performances in DL-based communication systems.
        Simulation results show that the high training SNR can improve the convergence performance in training, but it can also degrade the BLER performance in practical transmission.
      As a tradeoff, we introduce a training SNR set strategy, which shows trade-off between convergence and BLER performance.
      Furthermore, it is shown that training the autoencoder at low SNR can achieve BLER and MSE performance gains when the trained autoencoder is applied to high SNR scenario.
      These results provide a reliable design guidance to select the suitable training SNR and achieve optimal system performance.
\end{enumerate}
For potential applications, the DL-based autoencoder-represented communication system can be applied to complex channel conditions without a mathematically tractable model in, for examples, massive IoT ecosystems and high-speed Internet of Vehicles systems.

The remainder of this paper is organized as follows.
In Section \ref{sec2}, we describe the system model of a DL-based communication system.
Section \ref{sec3} presents an adaptive transmission scheme.
Section \ref{sec4} proposes the generalized data representation scheme for DL-based communication systems.
Section \ref{sec5} investigates the effect of SNR and analyzes the MSE performance of the autoencoder.
In Section \ref{sec6}, we present the numerical results of the proposed schemes and system performances.
Section \ref{sec7} concludes this paper.

%Throughout this paper, we use the following notations:
%$\mathbf A$ is a matrix; $\mathbf a$ is a vector; $a$ is a scalar; $\mathcal{A}$ is a set;
%$\|\mathbf A\|_F$ is the Frobenius norm of matrix $\mathbf A$;
%$[\mathbf A]_{i,:}$ is the $i$th row of matrix $\mathbf A$;
%$a_i$ is the $i$th element of vector $\textbf{a}$;
%$\mathbf 0$ is the zero vector; $\mathbf I$ is the identity matrix;
%$\mathbb{E}\{\cdot\}$ is used to denote expectation;
%$\lfloor\cdot\rfloor$ is the floor operation;
%$\binom{m}{n}$ denotes the number of combinations when choosing $n$ out of $m$.

\section{Deep Learning Based Communication Systems}
\label{sec2}
In this section, we describe the DL-based autoencoder for an end-to-end communication system, and then provide the research motivations of this paper.

\subsection{Autoencoder for End-to-End Communication Systems}
\begin{figure*}[t]
  \centering
  \epsfig{file=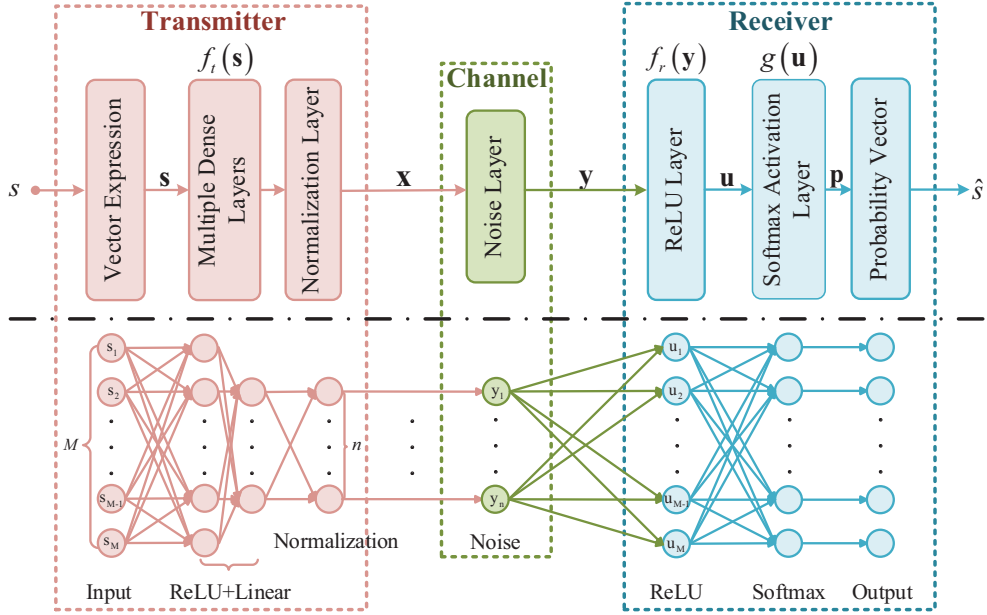,width=14cm}
  \caption{A DL-based communication system represented as an autoencoder with its NN structure \cite{8054694}.}
  \label{Sys}
\end{figure*}

\newcommand{\tabincell}[2]{\begin{tabular}{@{}#1@{}}#2\end{tabular}}
\begin{table}[h]
\caption{Activation Functions and Loss Functions}
\begin{center}
\begin{tabular}{c|c|c}
\hline
\multirow{5}{*}{\tabincell{c}{\textbf{Activation} \\\textbf{functions}}} &Linear     &$s_i$\\
\cline{2-3}
\multirow{5}{*}{}     &ReLU  &max$\{s_i, 0\}$\\
\cline{2-3}
\multirow{5}{*}{}     &Softmax    &$\frac{e^{u_i}}{\sum_{j=1}^Me^{u_j}}$\\
\cline{2-3}
\multirow{5}{*}{}     &Sigmoid    &$\frac{1}{1+e^{-u_i}}$\\
\cline{2-3}
\multirow{5}{*}{}     &tanh    &$\mathrm{tanh}(u_i)$\\
\hline
\multirow{2}{*}{\tabincell{c}{\textbf{Loss} \\\textbf{functions}}}      &MSE         &$\|\mathbf s-\mathbf p\|_2^2$\\
\cline{2-3}
\multirow{2}{*}{}        &Categorical cross-entropy  &$-\sum_{i=1}^M s_i\log(p_i)$\\
\hline
\end{tabular}
\label{function}
\end{center}
\end{table}

We consider a DL-based communication system represented as an autoencoder consisting of transmitter, channel, and receiver as shown in Fig. \ref{Sys}, where the corresponding NN structure is shown below.
The autoencoder describes a deep NN that applies unsupervised learning in order to reconstruct the input at the output \cite[Ch. 14]{Goodfellow-et-al-2016}.
At the transmitter, a message $s\in \{1, 2, \ldots, M\}$ is first transformed to a vector $\mathbf s\in \mathbb{R}^{M}$ after the vector expression processing, where, say, $M\in\{4, 8, 16, 32, 64\}$.
For example, if the message $s=2$ is transmitted, the corresponding vector expression is a one-hot vector $\mathbf s=[0,1,0,\ldots,0]^T$ in a conventional DL-based communication system.
Then, the multiple dense layers, including a rectified linear unit (ReLU) layer and a linear layer, apply the transformation $f_t: \mathbb{R}^{M}\mapsto\mathbb{R}^{n}$ to produce the transmitted signal for $n$ discrete channel uses \cite{8214233}.
The commonly used activation functions are shown in TABLE \ref{function}.
Finally, the normalization layer ensures the power constraint of the transmitted signal $\mathbf x=[x_1,\ldots,x_n]^T$ as $\mathbb{E}\{x_j^2\}\leq1$ ($j=1,\ldots,n$), where $\mathbb{E}\{\cdot\}$ denotes expectation.

The transmit channel is implemented by a noise layer with its output being the received signal $\mathbf y$ given by
\begin{IEEEeqnarray}{rCl}
\label{y}
\mathbf{y}=\mathbf x+\mathbf n
\end{IEEEeqnarray}
where $\mathbf{n}\sim\mathcal{N}(\mathbf{0},\sigma^2\mathbf{I}_n)$ denotes zero-mean additive white Gaussian noise (AWGN) vector where each element has variance $\sigma^2=(2RE_b/N_0)^{-1}$, and where $R$ is the data rate, $E_b$ is the energy per bit, and $N_0$ denotes the noise power spectral density.
Notably, there is no complex operation in the existing NN architectures, and the complex number is represented by two real numbers \cite{8054694}.
Consequently, we assume that all the channel coefficients have real values.
Furthermore, the autoencoder-represented communication system is suitable for any type of channel without a tractable mathematical model\footnote{We note that a real-world communication channel often does not have a tractable mathematical model.}.
That is to say, the autoencoder can be applied to any type of channel model as long as real datasets are available for training and learning.

At the receiver, the received signal $\mathbf y$ is passed through the ReLU layer\footnote{It can be shown by simulation, multiple ReLU layers do not improve the BLER performance for our problem.} to realize the transformation $f_r: \mathbb{R}^{n}\mapsto\mathbb{R}^{M}$.
The last layer of the receiver has a softmax activation as shown in TABLE \ref{function}, which is a generalization of the logistic function that compresses an $M$-dimensional vector of arbitrary real values to an $M$-dimensional probability vector $\mathbf p=[p_1,\ldots,p_M]^T$, where each element $p_i$ $(i=1, 2, \ldots, M)$ lies in the range (0, 1], and all the elements add up to one\footnote{We comment that the bit-to-bit vector representation (e.g., $00$, $01$, $10$, $11$) does not work in the autoencoder represented communication system.
The reasons is that the bit-to-bit vector representation has an all-zero vector (e.g., $00$) which cannot be compressed into a probability vector.} \cite{Goodfellow-et-al-2016}.
For the conventional autoencoder scheme, the estimated message $\hat{s}$ is obtained from the index of the element having the highest probability in $\mathbf p$.
Here, the BLER of DL-based communication systems is defined as
\begin{IEEEeqnarray}{rCl}
\label{BLER}
\mathrm{BLER}=\frac{1}{M}\sum_s\mathrm{Pr}(\hat{s}\neq s)\textrm{.}
\end{IEEEeqnarray}
Notably, the BLER equals the symbol error rate (SER) of the DL-based communication system.

The autoencoder based communication system can be trained offline using a large training dataset, while the iterative training process depends on the value of loss function in each iteration.
The most common loss functions are MSE and categorical cross-entropy as shown in TABLE \ref{function}, and these loss functions are determined by the vector expression $\mathbf s$ and the probability vector $\mathbf p$.
The training parameters of the autoencoder are produced to minimize the loss function.
Furthermore, the trained autoencoder with the fixed NN parameters is applied to practical communication scenarios online.

\subsection{Motivations}
The one-hot vector is the conventional data representation having only one non-zero element.
Thus, the data rate of the conventional DL-based communication system with one-hot vector is limited to
\begin{IEEEeqnarray}{rCl}
\label{Roh}
R_C=\frac{\log_2M}{n} \quad\textrm{bits/channel use}\textrm{.}
\end{IEEEeqnarray}
Over the last few years, the demand for high data rates has experienced unprecedented growth in communication systems \cite{6736746, 7414384}.
Therefore, providing a high data rate is essential for DL-based communication systems in future communications.

To improve the data rate, we propose two new autoencoder schemes:
\begin{enumerate}
  \item \emph{Adaptive transmission scheme.}
  For the conventional one-hot vector scheme, the DL-based autoencoder is trained over a fixed-size transmission vector with dimension $M$ at fixed SNR value, which can introduce two limitations.
  On one hand, the trained autoencoder for a certain value of $M$ cannot work in the scenarios with different values of $M$.
  On the other hand, the performance of DL-based communication systems is suboptimal when the trained autoencoder is applied to different SNR values.
  For these reasons, there is a need for a new transmission scheme for the autoencoder to improve the system performances, such as maximizing the data rate while satisfying the QoS constraint \cite{wu2013adaptive, 8241826}.
  Therefore, we propose an adaptive transmission scheme by adaptively selecting the optimal transmission vectors for different SNR values, where the optimization objective is to maximize the data rate with certain MSE constraint.
  \item \emph{Generalized data representation scheme.}
  From the definition of the data rate $R_{\emph{def}}=\frac{\textrm{Number of bits}}{\textrm{Channel uses}}$, it is obvious that, for the same channel environment, the data rate is proportional to the number of bits being conveyed.
  However, the size of transmission vector $M$ cannot be infinite due to the high complexity associated with deep NNs.
  Therefore, a new data representation scheme is required to meet the high data rate requirements in future communication systems.
  To address this issue, we design a generalized data representation scheme that employs a new vector structure instead of the one-hot vector.
  The new vector structure can be generalized and used for communication scenarios having different data rate requirements.
\end{enumerate}
Based on above discussions, we are motivated to develop data-rate driven transmission strategies for DL-based communication systems.

As for the system performances, the autoencoder that is trained offline using a fixed SNR value is expected to have robust performance for a wide SNR region online.
In \cite{8054694}, it was found that an unaccommodated training SNR will result in performance degradation of DL-based communication systems, but there is little theoretical analysis.
Consequently, the effect of the training SNR needs to be investigated and a reliable criterion needs to be developed for selecting training SNR values.
Furthermore, current literature on DL-based autoencoder research do not analyze its performance.
Therefore, we are motivated to develop an analytical framework to gain insights into the performance of DL-based communication systems.

\section{Adaptive Transmission Scheme}
\label{sec3}
In this section, an adaptive transmission scheme is employed in the DL-based communication system to maximize the data rate with the MSE constraint for different channel conditions.

\begin{figure*}[t]
  \centering
  \epsfig{file=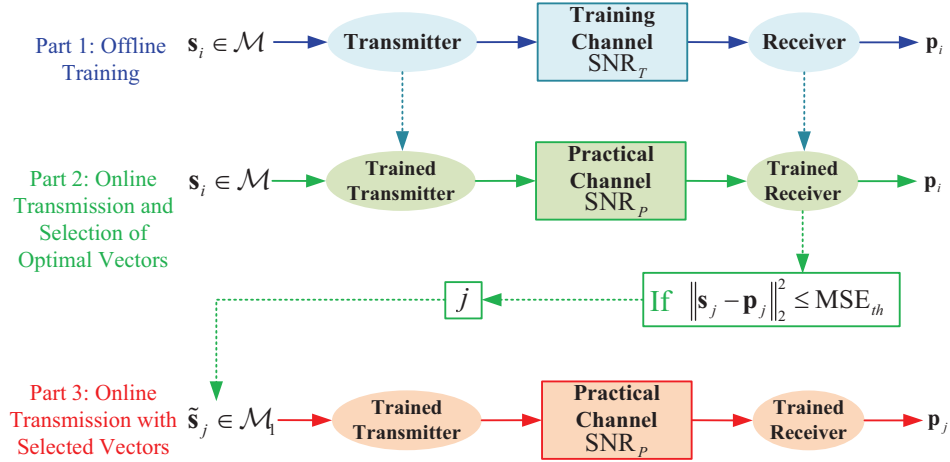,width=13cm}
  \caption{Adaptive transmission scheme applied to the DL-based communication system.}
  \label{Adasch}
\end{figure*}
Figure \ref{Adasch} shows the adaptive transmission scheme for the DL-based communication system, which consists of three parts.

The first part is offline training.
The autoencoder including the transmitter and receiver is trained offline using one-hot vectors $\mathbf{s}_1, \mathbf{s}_2,\ldots, \mathbf{s}_M$ over a fixed training SNR ($\mathrm {SNR}_T$), while $M$ should be suitably large\footnote{If $M$ is too large, the training complexity is prohibitive since the autoencoder must see every message at least once \cite{8054694}.}, for example $M=64$.
After training, the trained transmitter and receiver, which will be used in the second part and the third part, are produced with fixed parameters.

The second part is online transmission and selection of optimal vectors.
The second part includes three steps.
First, each one-hot vector $\mathbf{s}_i$ in set $\mathcal M=\{\mathbf s_1, \ldots,\mathbf s_M\}$ is transmitted through the trained transmitter/receiver once over the practical channel using an operating SNR value ($\mathrm{SNR}_P$).
Here, the receiver can obtain the probability vector $\mathbf{p}_i$ corresponding to the transmitted vector $\mathbf{s}_i$.
Second, the receiver calculates the MSE of each one-hot vector.
If the MSE of the $j$th vector ($\mathrm{MSE}_j$) is less than or equal to an MSE threshold, the receiver sends the label $j$ back to the transmitter.
In total, the receiver sends $M_1$ labels.
Third, according to the feedback labels, the transmitter forms a new vector set $\mathcal{M}_1$, which is defined as $\mathcal{M}_1=\{\tilde{\mathbf s}_j\}$, $j=1, \ldots, M_1$, where $\tilde{\mathbf s}_1, \tilde{\mathbf s}_2, \ldots, \tilde{\mathbf s}_{M_1}$ are the $M_1$ one-hot vectors selected from $\{\mathbf s_i\}$ with $M_1$ smallest MSE values.
The selection goal is to maximize the data rate and satisfy the MSE requirement as
\begin{IEEEeqnarray}{rCl}
\label{R1}
R_1=\,&&\max\frac{\log_2M_1}{n}\nonumber\\
&&\mathrm{s.t.}\;\:\:\|\mathbf s_j-\mathbf p_j\|_2^2\leq\mathrm{MSE}_{th}, \quad j=1, \ldots, M_1
\end{IEEEeqnarray}
where $M_1\leq M$ satisfying $M_1\in\{4,8,16,32,64\}$, and $\mathrm{MSE}_{th}$ is a preset MSE threshold.

The third part is online transmission with the selected vectors.
The selected $M_1$ one-hot vectors are used for the autoencoder online over the current channel with $\mathrm{SNR}_P$.

The main steps of the adaptive transmission scheme are summarized as follows:

\noindent\rule[0.12\baselineskip]{0.488\textwidth}{0.8pt}

    Steps of the Adaptive Transmission Scheme

\noindent\rule[0.12\baselineskip]{0.488\textwidth}{0.8pt}
\begin{enumerate}
      \item Train the autoencoder with a large training dataset consisting of all $M$ possible one-hot vectors offline.
      \item Each one-hot vector in $\mathcal{M}$ is transmitted through the trained autoencoder over the practical channel online.
      \item Calculate the practical MSE of each vector and select $\mathbf s_j$ according to (\ref{R1}).
      \item Feedback the label $j$ and form $\mathcal{M}_1=\{\tilde{\mathbf s}_j\}$.
      \item Encode the message symbol using $\mathcal{M}_1$ and transmit.
\end{enumerate}
\rule[0.12\baselineskip]{0.488\textwidth}{0.8pt}

\section{Generalized Data Representation Scheme}
\label{sec4}
In this section, we propose a generalized data representation scheme to improve the data rate for DL-based communication systems.

Instead of the conventional one-hot vector containing one non-zero entry, we consider a bit vector containing $m$ non-zero entries to improve the data rate for DL-based communication systems.
An $m$-order bit vector $\mathbf b\in \mathbb{R}^M$ is defined as
\begin{IEEEeqnarray}{rCl}
\label{b}
\mathbf{b}={\underbrace {[1\quad 0\quad \cdots\quad 0\quad 1\quad \cdots\quad 1\quad 0]}_{\textrm{$m$ 1's}}}^T
\end{IEEEeqnarray}
where $m=1,2,\cdots, \left\lfloor M/2\right\rfloor$ denotes the number of non-zero entries in $\mathbf b$, and $\lfloor\cdot\rfloor$ is the floor operation.
The bit vector provides $\binom{M}{m}$ possible messages for the transmission.
In general, the number of possible symbols in the constellation diagram is a power of 2.
For this reason, we only select $2^{\left\lfloor\log_2\binom{M}{m}\right\rfloor}$ out of $\binom{M}{m}$ possible symbols for communications.

Furthermore, for the autoencoder shown in Fig. \ref{Sys}, the vector $\mathbf s$ at the transmitter can be viewed as a probability distribution, and the probability vector $\mathbf p$ at the receiver is the corresponding estimated probability distribution.
The training goal of the autoencoder is to optimize $\mathbf p$ and reconstruct $\mathbf s$ while minimizing the loss function.

\begin{table}[t]
\caption{Results of messages transformed to vectors}
\begin{center}
\begin{tabular}{c|c|c}
\hline
\textbf{Message} & $16\times 1$ \textbf{One-hot Vector} & $8\times 1$ \textbf{GDR Vector}\\
\hline
1&  $[1,0,0,0,0,\ldots,0]^T$ & $[\frac{1}{2},\frac{1}{2},0,0,0,0,0,0]^T$\\
\hline
2&  $[0,1,0,0,0,\ldots,0]^T$ & $[\frac{1}{2},0,\frac{1}{2},0,0,0,0,0]^T$\\
\hline
3&  $[0,0,1,0,0,\ldots,0]^T$ & $[\frac{1}{2},0,0,\frac{1}{2},0,0,0,0]^T$\\
\hline
\vdots&  \vdots & \vdots\\
\hline
14&  $[0,\ldots,0,0,1,0,0]^T$ & $[0,0,\frac{1}{2},\frac{1}{2},0,0,0,0]^T$\\
\hline
15&  $[0,\ldots,0,0,0,1,0]^T$ & $[0,0,\frac{1}{2},0,\frac{1}{2},0,0,0]^T$\\
\hline
16&  $[0,\ldots,0,0,0,0,1]^T$ & $[0,0,\frac{1}{2},0,0,\frac{1}{2},0,0]^T$\\
\hline
\end{tabular}
\label{tabGDR}
\end{center}
\end{table}

Thus, motivated by the above discussions, we propose a generalized data representation as a probability distribution
\begin{IEEEeqnarray}{rCl}
\label{s}
\mathbf{s}={\underbrace {\left[\frac{1}{m} \quad 0\quad \cdots\quad 0\quad \frac{1}{m} \quad \cdots\quad \frac{1}{m} \quad 0\right]}_{\textrm{$m$ non-zero entries}}}^T
\end{IEEEeqnarray}
where the estimated message $\hat{s}$ can be obtained from the indices of elements with the $m$ highest probabilities in $\mathbf p$.
The conventional one-hot vector is a special case of the proposed GDR scheme when $m=1$.
Furthermore, the proposed GDR will be employed for the vector expression processing of the transmitter in Fig. \ref{Sys}.
As an example, when $M=16$, there are $16$ messages need to be transmitted.
For the conventional one-hot scheme, the corresponding vectors are $16$ different $16\times 1$ one-hot vectors $\mathbf s_i$, which are shown in the first column of TABLE \ref{tabGDR}.
For the proposed GDR scheme, the corresponding vectors are also $16$ different vectors, which can be $8\times 1$ GDR vectors with $m=2$.
The GDR scheme provides $\binom{M}{m}=28$ possible vectors for transmission, and we can randomly choose $16$ vectors\footnote{The vector selection is done here arbitrarily, and we leave the optimal vector selection as an open research problem.} as shown in the second column of TABLE \ref{tabGDR}.

The data rate of the DL-based communication system can be improved by employing the proposed GDR as
\begin{IEEEeqnarray}{rCl}
\label{R}
{R}=\frac{\left\lfloor\log_2\binom{M}{m}\right\rfloor}{n}\quad\textrm{bits/channel use}\textrm{.}
\end{IEEEeqnarray}
When $m=1$, the data rate is obtained for the conventional one-hot vector scheme in (\ref{Roh}).
The data rate increases with $m$, while the value of $M$ is suitably chosen and remains fixed.
The performance gain of the proposed GDR scheme will increase with vector size $M$.

The maximum achievable rate of the proposed GDR scheme in the DL-based communication system is derived as
\begin{IEEEeqnarray}{rCl}
\label{C}
{C}&&=\log_2(1+\mathrm{SNR})=\log_2\left(1+\frac{1}{\sigma^2}\right)\\
&&=\log_2\left(1+\frac{2E_b\cdot\left\lfloor\log_2\binom{M}{m}\right\rfloor}{N_0\cdot n}\right)\quad\textrm{bits/s/Hz}\nonumber \textrm{.}
\end{IEEEeqnarray}
It can be shown that the achievable rate can be improved by using the proposed GDR scheme in the DL-based communication system.
For example, when $M=16$, the proposed GDR scheme with $m=6$ has nearly $1.58$ (bits/s/Hz) performance gain compared with the conventional one-hot vector scheme at $E_b/N_0=20$ dB for seven channel uses.

Furthermore, the proposed GDR can be directly applied to the proposed adaptive transmission scheme by using the generalized data representation.
Combining the proposed two schemes, we obtain an adaptive GDR-based transmission scheme that can create further benefits for the DL-based communication system.

\section{Performance Analysis of the Autoencoder}
\label{sec5}
In this section, we provide a theoretical analysis of MSE performance for DL-based communication systems.
Such an analysis can be applied to two proposed schemes and other autoencoder-represented schemes.

\subsection{MSE Performance Analysis}
In Fig. \ref{Sys}, the output of the ReLU layer at receiver can be written as
\begin{IEEEeqnarray}{rCl}
\label{u}
\mathbf u&&=f_r(\mathbf y)\triangleq f_{ReLU}\left(\mathbf W_r\mathbf y+\mathbf b_r\right)
\end{IEEEeqnarray}
where $f_{ReLU}(a)=\max\{a, 0\}$; $\mathbf W_r$ and $\mathbf b_r$ denote the trainable parameters of the ReLU layer, and they are defined as
\begin{IEEEeqnarray}{rCl}
\label{Wrbr}
\mathbf W_r\!\!=\!\!
\left(\!\!\!\begin{array}{cccc}
w_{11} & w_{12} & \cdots & w_{1n} \\
w_{21} & w_{22} & \cdots & w_{2n} \\
\vdots & \vdots & \ddots & \vdots \\
w_{M1} & w_{M2} & \cdots & w_{Mn} \\
\end{array}\!\!\!\right)
\mathrm{and}\;
\mathbf b_r\!\!=\!\!
\left(\!\!\!\begin{array}{c}
b_1 \\b_2 \\ \vdots\\b_M\\
\end{array}\!\!\!\right)
\end{IEEEeqnarray}
respectively, where $w_{ij}$, $i=1, \ldots, M$, $j=1, \ldots, n$, represents the symmetric interaction term between unit $\mathrm{u_i}$ and unit $\mathrm{y_j}$ in Fig. \ref{Sys}, and $b_i$ is the bias term.
Thus, from (\ref{y}) and (\ref{u}), the $i$th element of $\mathbf u$ is given by
\begin{IEEEeqnarray}{rCl}
\label{ui}
u_i=\max \left\{[\mathbf W_r]_{i,:}(\mathbf x+\mathbf n)+b_i, 0\right\}
\end{IEEEeqnarray}
where $[\mathbf W_r]_{i,:}$ is the $i$th row of $\mathbf W_r$.

Next, a probability vector is derived from the softmax function at the receiver, and its $i$th element can be written as
\begin{IEEEeqnarray}{rCl}
\label{softmax}
p_i=\frac{e^{u_i}}{\sum_{k=1}^{M}e^{u_k}}\textrm{.}
\end{IEEEeqnarray}
From (\ref{ui})-(\ref{softmax}), in the offline training processing, different $\mathrm{SNR}=\frac{1}{\sigma^2}$ will lead to different trainable parameters $\mathbf W_r$ and $\mathbf b_r$, which will affect $u_i$ in (\ref{ui}).
As a result, $p_i$, the probability of the $i$th element is directly affected by the training SNR.
Also, in the online practical transmission, the trainable parameters $\mathbf W_r$ and $\mathbf b_r$ are constant since the autoencoder has been trained.
When the autoencoder is applied to a different SNR scenario online, it will lead to a different estimated probability vector $\mathbf p$ as well.
The effect of SNR will also be studied through simulations.

In Appendix A, it is shown that, based on (\ref{softmax}), the probability vector at the receiver in Fig. \ref{Sys} can be approximated as
\begin{IEEEeqnarray}{rCl}
\label{Papprox}
\mathbf p\approx\mathbf F\mathbf u
\end{IEEEeqnarray}
where $\mathbf F\in \mathbb R ^{M\times M}$ is a diagonal matrix that is equivalent to the effect of softmax activation layer.
It must be highlighted that, after training, the obtained $\mathbf F$ is constant when applying to online transmissions.

At the receiver, the output of the ReLU layer $\mathbf u$ consists of zero and non-zero elements as shown in \eqref{ui}.
In this paper, we aim to analyze the effect of SNR on MSE performance.
While the zero elements cannot reflect the characteristic of MSE, the non-zero output of the ReLU layer is considered and can be derived from (\ref{ui}) as
\begin{IEEEeqnarray}{rCl}
\label{unz}
\mathbf u_+=\mathbf W_r(\mathbf x+\mathbf n)+\mathbf b_r
\end{IEEEeqnarray}
if
\begin{IEEEeqnarray}{rCl}
\label{constrain}
[\mathbf W_r]_{i,:}(\mathbf x+\mathbf n)+b_i>0 \mathrm{.}
\end{IEEEeqnarray}
Thus, the probability vector $\mathbf p$ under the assumption of (\ref{constrain}) can be expressed as
\begin{IEEEeqnarray}{rCl}
\label{Pplus}
\mathbf p_+\approx\mathbf F_+\mathbf u_+
\end{IEEEeqnarray}
where $\mathbf F_+\in \mathbb R ^{M\times M}$ is the equivalent matrix of softmax activation layer in the non-zero case as \eqref{constrain}, and entries of $\mathbf F_+$ are fixed after training.

Here, the average MSE of the DL-based communication system in the case of (\ref{constrain}) can be given from (\ref{unz}) and (\ref{Pplus}) as
\begin{IEEEeqnarray}{rCl}
\label{MSEF}
\mathrm{MSE}&=&\mathbb{E}\left\{\|\mathbf p_+-\mathbf s\|_2^2\right\}\\
&\approx&\mathbb{E}\left\{\|\mathbf F_+(\mathbf W_r\mathbf x+\mathbf b_r)+\mathbf F_+\mathbf W_r\mathbf n-\mathbf s\|_2^2\right\}\nonumber\\
&=&\mathop{\mathbb{E}}\left\{\|\mathbf F_+(\mathbf W_r\mathbf x+\mathbf b_r)-\mathbf s\|_2^2\right\}+\|\mathbf F_+\mathbf W_r\|_2^2\sigma^2 \nonumber\mathrm .
\end{IEEEeqnarray}
After the autoencoder is trained over $\mathrm {SNR}_T$, the transformation parameters $\mathbf F_+$, $\mathbf W_r$ and $\mathbf b_r$ in (\ref{MSEF}) are constant, where $\sigma_{n_T}^2$ is the noise variance at the training scenario.
When the trained autoencoder is applied to the practical communication scenario with $\mathrm {SNR}_P$, the noise variance of the current practical channel scenario is $\sigma_{n_P}^2$.
For the non-zero case, it can be observed from (\ref{MSEF}) that, when $\sigma_{n_P}^2<\sigma_{n_T}^2$, the practical MSE performance will be better than that of the training scenario;
when $\sigma_{n_P}^2>\sigma_{n_T}^2$, the converse is true.
It indicates that the trained autoencoder can attain better system performance when it is applied to higher SNR scenario.
For the zero case in \eqref{ui}, the variance of noise has no effect on the MSE performance.
The MSE performance of the DL-based communication system will also be verified through simulations.

\subsection{Training SNR Set Strategy}
In conventional DL-based communication systems, the autoencoder is trained over a fixed SNR value offline, and it can suffer performance degradation when operating in environments having mismatched SNR values.
Here, we propose a training SNR set strategy by employing multiple training SNRs, and it will improve the diversity of training dataset to obtain robust performance.
For example, the training SNR set can be designed to $\mathcal{SNR_T} = \{-20, -10, 0, 10, 20\}$ dB for offline training.
Also, the system performance gain of the proposed training SNR set strategy will be shown by simulation results.

\section{Numerical Results}
\label{sec6}
\begin{table}[t]
\caption{Parameters for the autoencoder setup}
\begin{center}
\begin{tabular}{l|l}
\hline
\textbf{Parameter} & \textbf{Value} \\
\cline{1-2}
Optimizer&  Adam \cite{adam}\\
Loss function & MSE\footnotemark[7] \\
Epoch &  150 \\
Batch size & 45\\
Trained samples &2$\times10^4$\\
Test samples &1$\times 10^6$\\
\hline
\end{tabular}
\label{tab1}
\end{center}
\end{table}

\footnotetext[7]{For convenience, the MSE loss function is used to show the effect of SNR on the MSE performance and to verify the analysis in Subsection \ref{sec5}-A.}

In this section, we evaluate the numerical results of the proposed adaptive transmission scheme, the GDR scheme, the adaptive GDR-based transmission scheme, and the system performances in the DL-based communication system via simulations on the TensorFlow framework.
In all the simulations, the autoencoder is trained over the stochastic AWGN channel model with $n=7$ channel uses without exhaustive hyperparameter tuning.
Here, we use the same set of parameters for the autoencoder setup as described in TABLE \ref{tab1}.

\begin{table*}[t]
\caption{Training parameters of autoencoder}
\begin{center}
\begin{tabular}{c|c|c|c|c|c|c}
\hline
                       &\tabincell{c}{Vector \\size} &\tabincell{c}{Multiple \\dense layers} &\tabincell{c}{Normaliz-\\ation layer} &\tabincell{c}{ReLU \\layer} &\tabincell{c}{Softmax \\layer} &Total\\
\hline
\multirow{5}{*}{\tabincell{c}{Simulated \\parameters}} &$M=4$      &$55$  &$14$  &$32$  &$20$   &$121$\\
\cline{2-7}
\multirow{5}{*}{}     &$M=8$      &$135$  &$14$ &$64$  &$72$   &$285$\\
\cline{2-7}
\multirow{5}{*}{}     &$M=16$    &$391$   &$14$  &$128$  &$272$  &$805$\\
\cline{2-7}
\multirow{5}{*}{}     &$M=32$    &$1287$   &$14$  &$256$  &$1056$  &$2613$\\
\cline{2-7}
\multirow{5}{*}{}     &$M=64$    &$4615$   &$14$  &$512$  &$4160$  &$9301$\\
\hline
\tabincell{c}{Theoretical \\parameters} &$M$      &$(M+1)(M+n)$  &$2n$  &$M(n+1)$ &$M(M+1)$   &$(2M+3)(M+n)$\\
\hline
\end{tabular}
\label{tab2}
\end{center}
\end{table*}

TABLE \ref{tab2} presents the simulated and theoretical number of training parameters in autoencoder, where different size of the data representation $M$ is employed.
From TABLE \ref{tab2}, it is clear that the simulated number of trainable parameters increases with $M$ from $4$ to $64$, including the total number (of trainable parameters) and the number (of trainable parameters) in each layer except for the normalization layer.
The simulated results agree with the theoretical number of parameters as shown in the last row of TABLE \ref{tab2}.
The increasing number of training parameters leads to an increased complexity for training.
For the conventional one-hot vector, the data rate can be improved by increasing $M$ as shown in (\ref{Roh}) at the cost of high complexity.
While the data rate of the proposed GDR scheme can be improved by controlling the number of non-zero elements $m$ as well as the value of $M$ as shown in (\ref{R}).

\subsection{Performance of the Autoencoder and Conventional Communication System}
This subsection shows the simulated bit-error rate (BER) performance of the autoencoder scheme with one-hot vectors and the conventional communication scheme employing Hamming code, where the training SNR is $10$ dB.

\begin{figure}[t]
\begin{center}
\epsfig{file=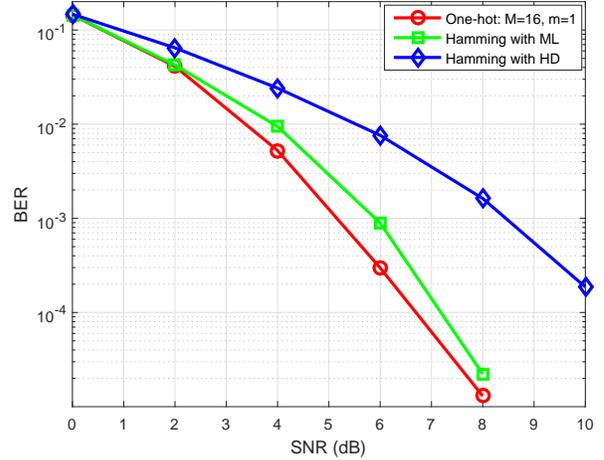,width=8.8cm}
\caption{Simulated BER performance for the autoencoder and conventional communication schemes.}
\label{BER_HM}
\end{center}
\end{figure}

Figure \ref{BER_HM} shows the simulated BER performance of the DL-based autoencoder scheme with $M=16$ and $m=1$ (one-hot vector) and the conventional communication scheme, where the conventional communication scheme employs binary phase-shift keying (BPSK) modulation and a ($7,4$) Hamming code with either binary hard-decision (HD) or maximum-likelihood (ML) decoding.
Given the same information transmission rate (transmitting four information bits over seven channel uses), it can be seen that the BER performance of the autoencoder scheme is better than that of the conventional communication scheme employing Hamming code with ML decoding or HD decoding.
It is worth pointing out that the autoencoder approach does not use any error control strategy for the noisy channel, and it still outperforms a classical scheme that employs error control strategy.
It was reported in \cite{8054694} that an autoencoder can achieve similar BLER performance compared to a conventional channel-coded scheme.

\begin{figure}[t]
\begin{center}
\epsfig{file=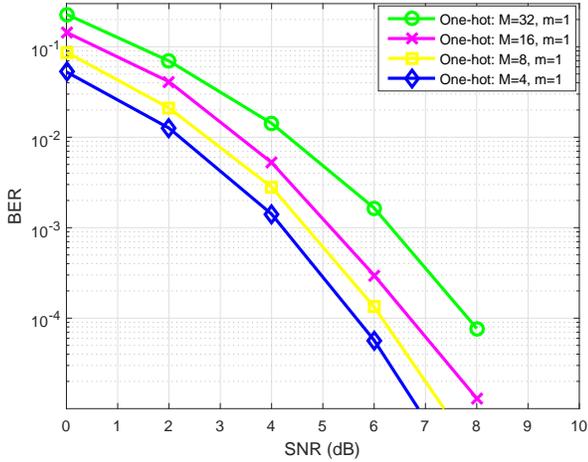,width=8.8cm}
\caption{Simulated BER for the autoencoder employing the conventional one-hot vector scheme with different vector size $M$.}
\label{Oneh_BER}
\end{center}
\end{figure}

Figure \ref{Oneh_BER} depicts the simulated BER performance of the DL-based autoencoder that employs gray coding and the conventional one-hot vector with the vector size $M=4,8,16,32$, where the training SNR is $10$ dB.
In Fig. \ref{Oneh_BER}, the BER of the conventional one-hot vector scheme increases when $M$ is varied from $4$ to $32$.

\subsection{Performance of the Proposed Adaptive Transmission Scheme}
In this subsection, we show the simulated BLER and MSE performance of the proposed adaptive transmission scheme in the DL-based communication system.
Here, the autoencoder is trained using $\mathrm {SNR}_T=5$ dB.

\begin{figure}[t]
\begin{center}
\epsfig{file=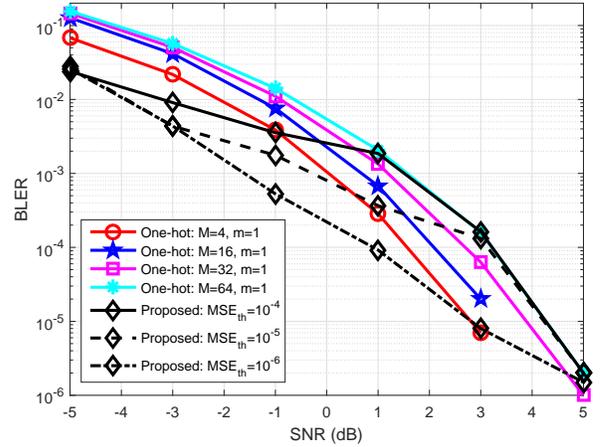,width=8.8cm}
\caption{Simulated BLER for the autoencoder with conventional one-hot vector and proposed adaptive transmission schemes, while the MSE thresholds are $10^{-4}$, $10^{-5}$, and $10^{-6}$.}
\label{CABLER}
\end{center}
\end{figure}

\begin{table*}[t]
\caption{The number of adaptively selected vectors $M_1$ for different SNR values and $\mathrm{MSE}_{th}$}
\begin{center}
\begin{tabular}{c|c|c|c|c|c|c}
\hline
$M_1$    &SNR=$-5$ dB      &SNR=$-3$ dB  &SNR=$-1$ dB &SNR=$1$ dB  &SNR=$3$ dB   &SNR=$5$ dB\\
\hline
$\mathrm{MSE}_{th}=10^{-4}$     &$4$    &$16$   &$32$  &$64$  &$64$  &$64$\\
\hline
$\mathrm{MSE}_{th}=10^{-5}$     &$4$    &$4$   &$16$  &$32$  &$64$  &$64$\\
\hline
$\mathrm{MSE}_{th}=10^{-6}$     &$4$    &$4$   &$4$  &$16$  &$32$  &$64$\\
\hline
\end{tabular}
\label{tab3}
\end{center}
\end{table*}

Figure \ref{CABLER} depicts the simulated BLER performance of the DL-based autoencoder that employs the proposed adaptive transmission scheme and the conventional one-hot vector scheme, where the MSE thresholds are $10^{-4}$, $10^{-5}$, and $10^{-6}$.
First, it can be seen from Fig. \ref{CABLER} that the BLER of the conventional one-hot vector scheme increases when $M$ is varied from $4$ to $64$, since smaller value of $M$ requires less trainable parameters as shown in TABLE \ref{tab2}.
With the same training dataset, the less trainable parameters contribute to better training accuracy.
Second, for the proposed adaptive transmission scheme, the BLER increases when the MSE threshold is increased from $10^{-6}$ to $10^{-4}$ in Fig. \ref{CABLER}.
The reason is that, to maximize the data rate, a lower MSE threshold (means the tighter bound) requires smaller $M_1$ to satisfy the MSE constraint, which results in lower BLER.
As shown in TABLE \ref{tab3}, for each MSE threshold, the number of selected vectors $M_1$ adaptively increases from $4$ to $64$ with the increasing SNR value.
For example, for $\mathrm{MSE}_{th}=10^{-5}$, when SNR is changing from $-5$ dB to $5$ dB, the $M_1$ value changes accordingly as $4, 4, 16, 32, 64, 64$.
For this reason, higher SNR value makes it easy to meet the MSE requirement and as a result, a larger value $M_1$ is obtained for maximizing the data rate.
Fig. \ref{CABLER} shows that, when the data rates are the same, i.e. $M=M_1$, the adaptive transmission scheme can reduce the BLER of the one-hot vector scheme by $80\%$.
The reason for the performance gain is that the proposed adaptive transmission scheme can select the optimal vectors that meet the MSE requirement as shown in (\ref{R1}).

\begin{figure}[t]
\begin{center}
\epsfig{file=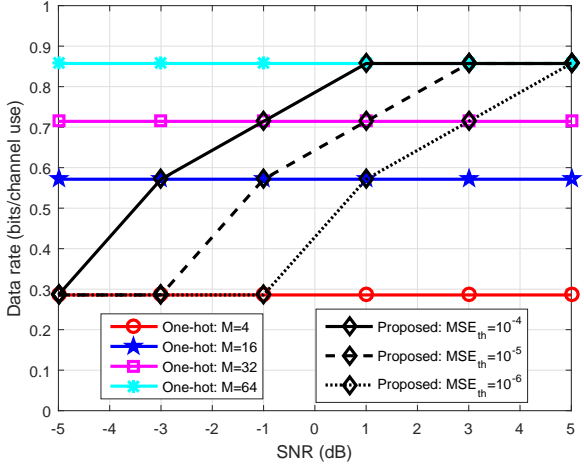,width=8.8cm}
\caption{Data rate performance for the autoencoder with conventional and adaptive transmission schemes, while the MSE thresholds are $10^{-4}$, $10^{-5}$, and $10^{-6}$.}
\label{Datarate}
\end{center}
\end{figure}

Figure \ref{Datarate} illustrates the data rate performance of the autoencoder that employs the conventional one-hot vector scheme and the proposed adaptive transmission scheme with the MSE thresholds being $10^{-4}$, $10^{-5}$, and $10^{-6}$.
From Fig. \ref{Datarate}, we observe that the data rates of the conventional one-hot vector scheme are constant for all SNR values.
However, in Fig. \ref{Datarate}, the data rate of the proposed adaptive transmission scheme increases with SNR as shown in TABLE \ref{tab3}.
From Fig. \ref{CABLER} and Fig. \ref{Datarate}, it can be seen that the proposed adaptive transmission scheme can obtain better BLER performance than that of the conventional one-hot vector scheme when operating at the same data rate.

\begin{figure}[t]
\begin{center}
\epsfig{file=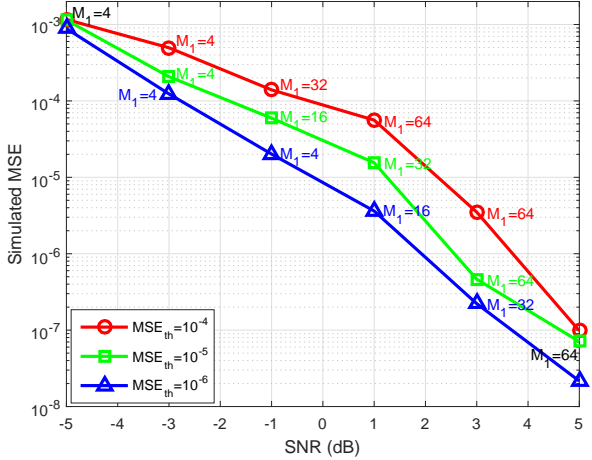,width=8.8cm}
\caption{Simulated MSE for the autoencoder employing the adaptive transmission scheme with the MSE thresholds being $10^{-4}$, $10^{-5}$ and $10^{-6}$.}
\label{MSE}
\end{center}
\end{figure}

Figure \ref{MSE} presents the simulated MSE performance for a practical communication system that employs the proposed adaptive transmission scheme with MSE thresholds being $10^{-4}$, $10^{-5}$ and $10^{-6}$.
It is seen from Fig. \ref{MSE} that the simulated MSE of the proposed adaptive transmission scheme increases with MSE threshold.
Furthermore, the simulated MSE of the proposed scheme decreases while the SNR increases, which is consistent with the prediction in (\ref{MSEF}).
As expected, when the simulated MSE reaches the corresponding MSE threshold, the number of selected vectors $M_1$ is almost $64$ which is the maximum value, and the maximum data rate is obtained.

\subsection{Performance of the Proposed GDR Scheme}
This subsection shows the BLER performance and the maximum achievable rate of the proposed GDR scheme in the DL-based communication system, where the training SNR is $5$ dB.

\begin{figure}[t]
\begin{center}
\epsfig{file=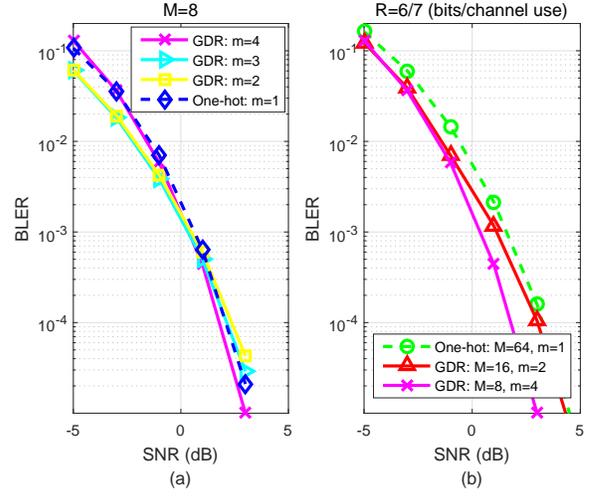,width=8.8cm}
\caption{Simulated BLER for the autoencoder employing different data representations with (a) $M=8$ and (b) $R=6/7$ (bits/channel use), while the trained SNR is 5 dB.}
\label{BLER1}
\end{center}
\end{figure}

\begin{table*}[t]
\caption{Data rate of the DL-based communication system}
\begin{center}
\begin{tabular}{c|c|c|c|c|c|c}
  \hline
   & One-hot & GDR & GDR & GDR & GDR & One-hot \\
   & $M\!\!=\!\!8,m\!\!=\!\!1$ & $M\!\!=\!\!8, m\!\!=\!\!2$ & $M\!\!=\!\!8, m\!\!=\!\!3$ & $M\!\!=\!\!8, m\!\!=\!\!4$ & $M\!\!=\!\!16, m\!\!=\!\!2$ & $M\!\!=\!\!64, m\!\!=\!\!1$ \\
  \hline
   \tabincell{c}{Data rate\\ (bits/channel use)} & $3/7$ & $4/7$ & $5/7$ & $6/7$ & $6/7$ & $6/7$  \\
  \hline
\end{tabular}
\label{tab4}
\end{center}
\end{table*}

Figure \ref{BLER1} shows the simulated BLER performance of the DL-based communication system that employs the proposed GDR and conventional one-hot vector schemes, while the schemes in (a) have the same vector size $M=8$ and the schemes in (b) have the same data rate $R=6/7$ (bits/channel use).
In Fig. \ref{BLER1} (a), for the same vector size $M=8$, the proposed GDR schemes ($m=2, 3, 4$) obtain comparable BLER performances when compared to the conventional one-hot vector scheme ($m=1$).
It indicates that, with the same vector size, the number of non-zero elements in $\mathbf s$ has little effect on the BLER performance.
Even the BLER performances are similar, the data rates of the GDR schemes and the one-hot vector scheme are different and they are shown in TABLE \ref{tab4}.
It can be seen from TABLE \ref{tab4} that, with $M=8$, the data rates of the proposed GDR schemes are $R=6/7, 5/7, 4/7$ (bits/channel use) respectively with $m=4, 3, 2$.
The data rates of all GDR schemes are greater than that of the conventional one-hot vector scheme as $R=3/7$ (bits/channel use), and the GDR scheme with $m=4$ can double the data rate of the one-hot vector scheme.
In Fig. \ref{BLER1} (b), with the same data rate $R=6/7$ (bits/channel use) including the proposed schemes $M=8$ with $m=4$, $M=16$ with $m=2$, and the conventional scheme $M=64$ with $m=1$, the proposed GDR schemes have better BLER performance than that of the conventional one-hot vector scheme, and these performance gains are achieved with the GDR schemes with lower training complexity, i.e., less number of training parameters as shown in TABLE \ref{tab2}.
Obviously, the BLER decreases with the vector size $M$  for the same reason as that in Fig. \ref{CABLER}.
Furthermore, it can be found that the proposed GDR scheme can avoid the performance degradation by increasing $m$, when the transmission message size is large.
For example, the proposed GDR scheme $M=16$ with $m=8$ can transmit $2^{\lfloor \log_2\binom{16}{8}\rfloor}=32768$ messages by using $16\times 1$ vectors.
However, to achieve the same data rate, the size of one-hot vector should be $32768\times 1$ at least, which will lead to significant BLER performance degradation.
In both Fig. \ref{BLER1} (a) and (b), the simulated BLER is less than $10^{-5}$ when the SNR is 5 dB, which demonstrates that the autoencoder attains a high accuracy with sufficient training over $\mathrm {SNR}_T=5$ dB.

\begin{figure}[t]
\begin{center}
\epsfig{file=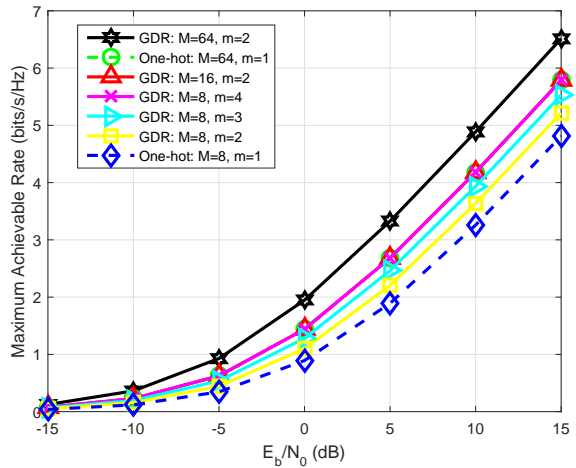,width=8.8cm}
\caption{Maximum achievable rate for the autoencoder with different data representations.}
\label{MSP}
\end{center}
\end{figure}

Figure \ref{MSP} illustrates the maximum achievable rate of a DL-based communication system employing different data representations.
It can be seen from Fig. \ref{MSP} that, with $M=8$, the maximum achievable rate increases when the order $m$ increases from $1$ to $4$, which is consistent with the result in (\ref{C}).
This shows that the proposed GDR scheme can obtain a remarkable achievable rate improvement.
Notably, the performance gain of the proposed GDR scheme is increased when the vector size $M$ increases.
As shown in Fig. \ref{MSP}, the GDR scheme employing $M=64$ with $m=2$ has a great performance gain when compared with the conventional scheme employing $M=64$ with $m=1$.
Besides, the maximum achievable rate of the proposed GDR schemes ($M=8$ with $m=4$ and $M=16$ with $m=2$) is same as that of the conventional one-hot vector scheme ($M=64$, $m=1$) in Fig. \ref{MSP}.
To obtain the same achievable rate with the GDR scheme, the conventional one-hot vector scheme needs to increase the vector size $M$, which has been shown in Fig. \ref{BLER1} to degrade the BLER performance.

\begin{figure}[t]
\begin{center}
\epsfig{file=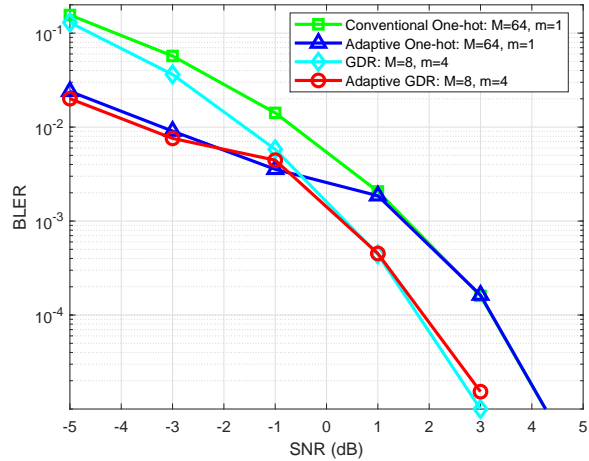,width=8.8cm}
\caption{Simulated BLER for the autoencoder employing four different schemes, while the trained SNR is 5 dB.}
\label{BLERall}
\end{center}
\end{figure}

\footnotetext[8]{Except the adaptive GDR-based transmission scheme, the BLER performance results of the other three schemes have been shown in Fig. \ref{CABLER} and Fig. \ref{BLER1} (b).}

Figure \ref{BLERall} presents the simulated BLER performance of the DL-based communication system that employs four schemes, including the conventional one-hot vector ($M=64$, $m=1$), the adaptive transmission (based on one-hot vector), the GDR ($M=8$, $m=4$), and the adaptive GDR-based transmission schemes\footnotemark[8].
The MSE threshold of the two adaptive schemes is $\mathrm{MSE}_{th}=10^{-4}$.
The achievable maximal data rate of all four schemes is the same, $R=6/7$ (bits/channel use).
In Fig. \ref{BLERall}, the adaptive GDR-based transmission scheme achieves the best BLER performance in all SNR regions, i.e., the adaptive GDR-based transmission scheme outperforms the GDR ($M=8$, $m=4$) scheme in low SNR region and the adaptive transmission (based on one-hot vector) scheme in high SNR region.
With the $\mathrm{MSE}_{th}=10^{-4}$, the adaptive GDR-based transmission scheme selects $M_1$ vectors for transmission, when SNR is changing from $-5$ dB to $5$ dB, the $M_1$ value changes accordingly as $8, 16, 32, 64, 64, 64$.
It can be seen that, when $\mathrm{SNR}=-5, -3, -1$ dB, the BLER of the adaptive GDR-based transmission scheme is similar to that of the adaptive transmission (based on one-hot vector) scheme since they have similar $M_1$;
when $\mathrm{SNR}=1, 3, 5$ dB, the BLER of the adaptive GDR-based transmission scheme is similar to that of the GDR scheme ($M=8$, $m=4$), and the reason is that the adaptive GDR-based transmission scheme selects all $64$ vectors for usage, in this case, the adaptive GDR-based transmission scheme is equal to the GDR scheme ($M=8$, $m=4$).

\subsection{Performance Comparison of Different Training SNR}

\begin{figure}[t]
\begin{center}
\epsfig{file=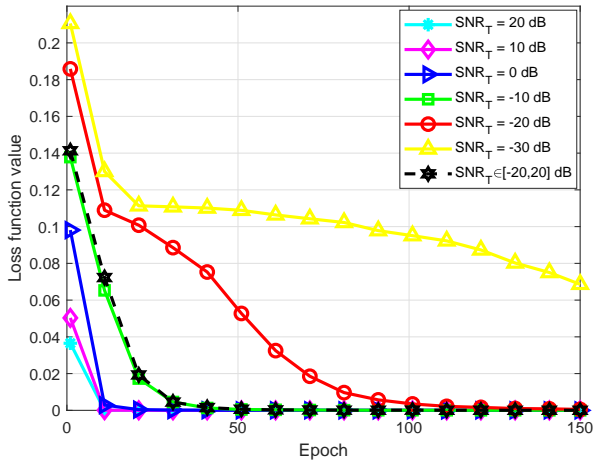,width=8.8cm}
\caption{Simulated loss function performance of autoencoder in training process, while different fixed training SNRs and training SNR set are employed.}
\label{TSNR}
\end{center}
\end{figure}

In this subsection, we investigate the effect of training SNR on system performance including the loss function performance in training process, the simulated BLER, and MSE performances in practical transmission process.
Here, the data representation parameters are $M=8 $ and $m=1$, and $\mathrm{SNR}_T$ denotes the training SNR.

Figure \ref{TSNR} shows the simulated loss function performance in training processing, when the autoencoder is trained over different SNRs and SNR set.
The SNR set is designed as $\mathcal{SNR_T}=\{-20, -10, 0, 10, 20\}$ dB, which includes all the fixed SNRs except for $-30$ dB.
In Fig. \ref{TSNR}, an epoch is the process that the entire training dataset is passed through the autoencoder once.
As shown in Fig. \ref{TSNR}, when the $\mathrm{SNR}_T$ is increased from $-20$ dB to $20$ dB, the loss function value decreases and the convergence of loss function improves, which indicates that the good channel environment contributes to the improvement of the training performance.
However, with $\mathrm{SNR}_T=-30$ dB, the loss value does not converge within $150$ epoches.
Furthermore, it can be seen from Fig. \ref{TSNR} that, the loss value of the autoencoder training with SNR set is similar to that of the autoencoder training with $\mathrm{SNR}_T=-10$ dB.
The simulated results suggest that the training SNR has significant effect on the training performance of the autoencoder.

\begin{figure}[t]
\begin{center}
\epsfig{file=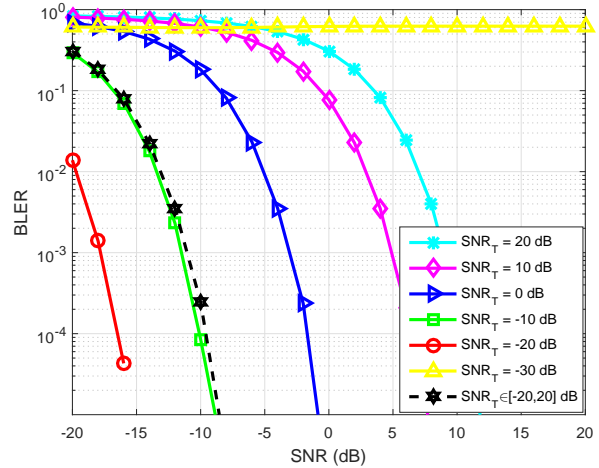,width=8.8cm}
\caption{Simulated BLER for the DL-based communication system employing the trained autoencoder with different fixed training SNRs and training SNR set.}
\label{TSNRBLER}
\end{center}
\end{figure}

Figure \ref{TSNRBLER} depicts the simulated BLER performance of the practical DL-based communication system employing the trained autoencoder with different fixed training SNRs and training SNR set.
In Fig. \ref{TSNRBLER}, the BLER decreases with $\mathrm{SNR}_T$ ranging from $20$ dB to $-20$ dB.
The reason is that, with the lower training SNR (that is to say the worse channel environment), the autoencoder needs to learn more features to reconstruct the input at the output, which leads to a robust autoencoder and better BLER performance.
However, the training SNR has a lower bound for the autoencoder.
As shown in Fig. \ref{TSNRBLER}, when $\mathrm{SNR}_T=-30$ dB, the BLER is approximately 0.6, which demonstrates that the autoencoder trained over this channel environment cannot learn the features anymore.
It is consistent with the non-convergence performance of the loss function with $\mathrm{SNR}_T = -30$ dB in Fig. \ref{TSNR}.
Besides, Fig. \ref{TSNRBLER} shows that the BLER performance of the training SNR set scheme is similar to that of $\mathrm{SNR}_T = -10$ dB scheme, which is almost the best performance except for the $\mathrm{SNR}_T =-20$ dB scheme.
It shows that training with SNR set can improve the generalization performance of the autoencoder.
From Fig. \ref{TSNR} and Fig. \ref{TSNRBLER}, it can be found that, with a higher training SNR value, the autoencoder obtains better convergence performance in training but worse BLER performance.
The simulated results indicate that the training SNR will directly affect the system performance, which agrees with the analysis in Subsection V-A.

\begin{figure}[t]
\begin{center}
\epsfig{file=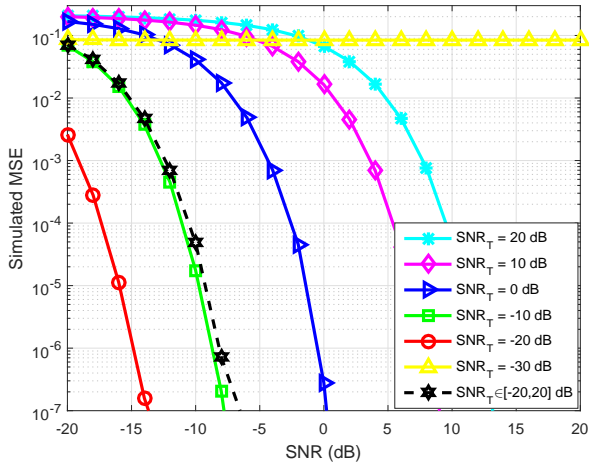,width=8.8cm}
\caption{Simulated MSE for the DL-based communication system employing the trained autoencoder with different fixed training SNRs and training SNR set.}
\label{TMSE}
\end{center}
\end{figure}

Figure \ref{TMSE} illustrates the simulated MSE performance of the practical DL-based communication system employing different trained autoencoders, while the training SNRs include different fixed SNRs and SNR set.
In Fig. \ref{TMSE}, it is clear that the MSE decreases when SNR is increased.
It indicates that the MSE performance improves when the trained autoencoder is applied to a higher SNR scenario, which is consistent with the analysis in (\ref{MSEF}).
Furthermore, the simulated MSE performance in Fig. \ref{TMSE} is similar to the BLER performance as shown in Fig. \ref{TSNRBLER} for the same reasons.

\section{Conclusion and Future Research}
\label{sec7}
In this paper, we proposed two new transmission schemes to address the problem of limited data rate in DL-based communication system using autoencoder.
We designed an adaptive transmission scheme for different channel conditions to maximize the data rate with a mean square error constraint.
Furthermore, we proposed the GDR scheme to obtain higher data rate than the conventional one-hot vector scheme with a similar BLER performance.
Besides, the effect of training SNR and MSE performance were analyzed and verified by simulations.
We discovered that high training SNR can lead to good convergence in training process but worse BLER performance for practical transmission.
We also introduced a training SNR set strategy to address the tradeoff between convergence and error rate.
It was shown that the autoencoder trained over a low SNR can attain better BLER and MSE performances when operating in the high SNR region.
As a result, it is concluded that training the autoencoder at a lower SNR value, in general, will lead to good system performance.

For a low SNR value, say, $\mathrm{SNR}_T=-30$ dB, numerical results indicate that the loss function value does not converge and the BLER degrades dramatically.
This suggests that the studied autoencoder system is unable to learn from very noisy data set.
It should be emphasized that the current system assumes neither knowledge about the noise nor about the system model.
Therefore, one interesting research problem is to study the low SNR communication using DL techniques when partial knowledge about the noise and the system model is known.

To further improve the performance of the DL-based communication systems, we can possibly employ the ensemble method where the results of a set of individual NNs are combined to estimate the transmitted message.
In classifier problems, it has been shown that the ensemble method is effective in improving accuracy or decomposing a complex problem into easier subproblems \cite{krawczyk2017ensemble}.

\appendices
\section{ Derivation of (\ref{Papprox}) }
Let
\begin{IEEEeqnarray}{rCl}
\label{Pdecom}
\mathbf p=\mathbf F\mathbf u \mathrm .
\end{IEEEeqnarray}
According to (\ref{softmax}), eq. (\ref{Pdecom}) can be formulated as
\begin{IEEEeqnarray}{rCl}
\label{Punfold}
\frac{1}{\sum_k^{M}e^{u_k}}\!\!
\left(\!\!\!\begin{array}{c}
e^{u_1}\\e^{u_2} \\ \vdots\\e^{u_M}\\
\end{array}\!\!\!\right)
\!\!=\!\!
\left(\!\!\!\begin{array}{cccc}
f_{11} & f_{12} & \cdots & f_{1M} \\
f_{21} & f_{22} & \cdots & f_{2M} \\
\vdots & \vdots & \ddots & \vdots \\
f_{M1} & f_{M2} & \cdots & f_{MM} \\
\end{array}\!\!\!\right)\!\!\!
\left(\!\!\!\begin{array}{c}
u_1 \\u_2 \\ \vdots\\u_M\\
\end{array}\!\!\!\right)
\end{IEEEeqnarray}
and we can obtain that
\begin{IEEEeqnarray}{rCl}
\label{eui}
e^{u_i}\!\!=\!\!\left(f_{i1}u_1\!+\!f_{i2}u_2\!+\!\cdots\!+\!f_{ii}u_i\!+\!\cdots\!+\!f_{iM}u_M\right)\!\!\sum_{k=1}^{M}e^{u_k}\mathrm .
\end{IEEEeqnarray}

Next, $e^{u_i}$ can be approximated according to the Taylor's theorem as
\begin{IEEEeqnarray}{rCl}
\label{exTaylor}
e^{u_i}\approx 1+u_i+\frac{u_i^2}{2!}+\cdots+\frac{u_i^N}{N!}
\end{IEEEeqnarray}
where $N$ is a sufficiently large integer.

Finally, combining (\ref{eui}) and (\ref{exTaylor}), we can derive the elements of matrix $\mathbf F$ as
\begin{IEEEeqnarray}{rCl}
\label{fij}
f_{ij}\!\!\approx\!\!\left\{\begin{array}{ll}\frac{1}{\sum_{k=1}^{M}e^{u_k}}\left(u_i^{-1}\!+\!1\!+\!\frac{u_i}{2!}\!+\!\cdots\!+\!\frac{u_i^{N-1}}{N!}\right) \;i\!=\!j\\
0 \qquad i\neq j
\end{array}\right.
\mathrm .
\end{IEEEeqnarray}

Thus, eq. (\ref{fij}) shows that the probability vector $\mathbf p$ at the receiver can be approximated as (\ref{Papprox}).

\section*{Acknowledgment}
We thank all anonymous reviewers and the editor for their constructive comments that have significantly improve the original manuscript.
In particular, we thank the editor Prof. Vaneet Aggarwal for suggesting us investigate the adaptive GDR-based transmission scheme.

\bibliographystyle{IEEEtran}
\bibliography{IEEEabrv,DL}

\newpage

\section*{\textbf{Corrections to ``Data-Rate Driven Transmission Strategies for Deep Learning Based Communication Systems''}}
In \cite{8964474}, the simulation results are obtained by using the BatchNormalization in the TensorFlow framework.
The BatchNormalization was used to ensure the power constraint of the transmitted signal $\mathbf x=[x_1,\ldots,x_n]^T$ as $\mathbb{E}\{x_j^2\}\leq1$.
However, in the testing phase, it was found the trainable parameters of the BatchNormalization layer were not updated from the training results, which led to unbounded transmission power.
Thus, the results in Fig. 8 (a) and Fig. 12 are incorrect, because they did not satisfy the converse bounds of average power constraint over the additive white Gaussian noise channel \cite{7303958}.

With correct normalization (e.g. l2 normalization), Fig. 8 (a) and Fig. 12 in \cite{8964474} should appear as Fig. \ref{BLER1} and Fig. \ref{TSNRBLER}, respectively.
\begin{figure}[ht]
\begin{center}
\epsfig{file=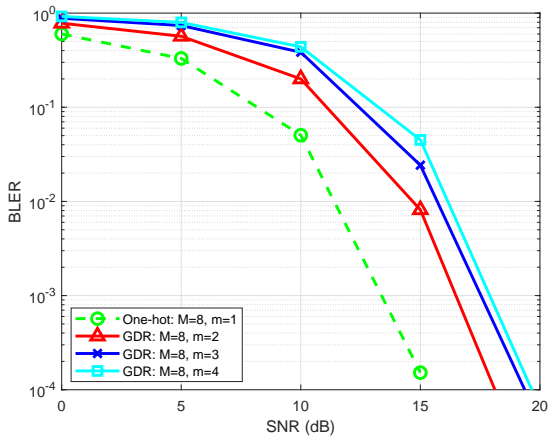,width=8.2cm}
\caption{Simulated BLER for the autoencoder employing different data representations, when the training SNR is 10 dB.}
\label{BLER1}
\end{center}
\end{figure}

\begin{figure}[ht]
\begin{center}
\epsfig{file=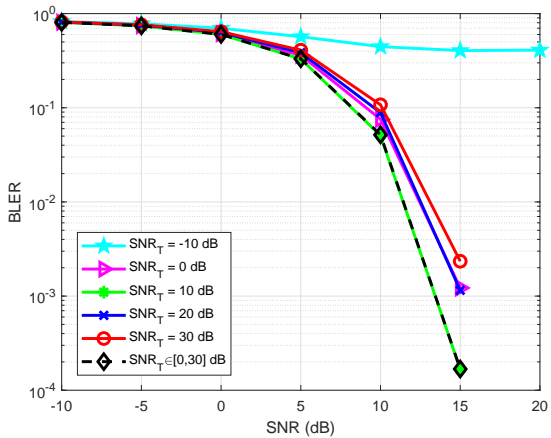,width=8.2cm}
\caption{Simulated BLER for the communication system employing the trained autoencoder having different fixed training SNRs and training SNR set.}
\label{TSNRBLER}
\end{center}
\end{figure}

In Fig. \ref{BLER1}, the proposed GDR schemes ($m=2, 3, 4$) obtain slightly worse BLER performances than the conventional one-hot vector scheme ($m=1$).
However, the data rates of the proposed GDR schemes are $R=6/7, 5/7, 4/7$ (bits/channel use) respectively with $m=4, 3, 2$, which are greater than that of the conventional one-hot vector scheme as $R=3/7$ (bits/channel use).

In Fig. \ref{TSNRBLER}, the BLER performances are comparable while using the fixed training SNRs ($\mathrm{SNR}_T=0,10,20,30$ dB) and training SNR set ($\mathcal{SNR_T}=\{0, 10, 20, 30\}$ dB).
However, when $\mathrm{SNR}_T=-10$ dB, the BLER is approximately $0.4$, which demonstrates that the autoencoder trained over this channel environment cannot learn the features anymore.
It can be shown that, with a higher training SNR value, the autoencoder achieves better BLER performance.

The source codes used for generating the results in this work are available for the public \cite{codes}.

\section*{Acknowledgment}
We thank the editor-in-chief Prof. Tolga Mete Duman and his student Berke Eren for their constructive comments that have helped the authors identify and correct this error.

\end{document}